\documentstyle[11pt]{article}
\newcommand{\be}{\begin{eqnarray}}

\newcommand{\ee}{\end{eqnarray}}

\title{
	\begin{flushright}
	{\normalsize TPI-MINN-93/51-T \\
	NUC-MINN-93/27-T \\
        UMN-TH-1223/93 \\
	October 1993 \\}
	\end{flushright}
\bf B+L Non-Conservation as a Semi-Classical Process \footnote{
Presented at the XXXIII Crakow School of Physics, Zakopane, Poland
June 1-12 1993}}
\author{
	Larry McLerran \\
	{\small\it Theoretical Physics Institute,}  \\
        {\small\it  School of Physics and Astronomy,} \\
 	{\small\it University of Minnesota, Minneapolis, MN 55455} \\
	 }
\date{}

\begin{document}

\maketitle

\begin{center}
{\bf Abstract}\\
\end{center}
I discuss the problem of computing B+L violation at high energies as that
of solving classical differential equations.  These equations
involve boundary conditions at initial and a final times which
are anti-Feynman,
and involve solving non-linear differential equations
which are complex, even if the original Hamiltonian was real.

\vfill \eject

\section{Introduction}

Electroweak theory was constructed so that the symmetries of the
classical equations of motion require that the classical baryon plus
lepton number current is conserved,
\be
	\partial_\mu J^\mu_{B+L} = 0
\ee
The theory is constructed in this way
because of the very long lifetime of the proton $\tau_p \ge 10^{32}~yr$
If baryon number was violated at the level of the classical equations of
motion, we would naively expect a proton lifetime of the order of a
typical weak decay time.  The limits on the proton lifetime
on the other hand naively require that the scale of $B+L$ violation in
elementary processes be larger than about $10^{15}~ GeV$.

On the other hand, it has been known for some time that
in first order in quantum corrections, $B+L$ is no longer
conserved\cite{adler}
and
\be
	\partial_\mu J^\mu_{B+L} = 2 N_{fam}~ \partial_\mu K^\mu_{CS}
\ee
where the Chern-Simons current is
given as
\be
	\partial_\mu K^\mu = {{\alpha_W} \over {8\pi}} FF^d
\ee
where $N_{fam}$ is the number of quark families, $F$ is the electroweak
field strength and $F^d_{\mu \nu} = {1 \over 2 } \epsilon_{\mu \nu \sigma \rho}
F^{\sigma \rho} $.  We will work in this paper in the limit where
the Weinberg angle is zero, that is, we will ignore the effects
of the $U(1)$ field of $ SU(2) \times U(1)$

We can understand the factor of $N_{fam}$ in the above expression since
the electroweak Lagrangian is blind to family, and therefore
all families must be produced equally.  In the same way, the theory
is blind to the difference between quark and lepton so the the
process must produce three quarks and 1 lepton for each family,
that is the change in baryon number must equal that of lepton
number.  Electroweak theory however know the difference between upper
and lower components of a doublet and therefore the minimal process
need and in fact does involve only one particle of the two
in each doublet.

By the anomaly equation, we see that the only way baryon number can be
changed is by having a non-zero value of $\int d^4x~ FF^d$.  If we
formulate the computation of physical quantities as a Euclidean path
integral, in order to generate a baryon number violating amplitude,
we must have contributions from the sector of the theory where
this integral is non-zero.  Moreover, the anomaly equation involves
an integer change in Chern-Simons charge.  One can prove that the
topological charge change is in fact integer, and the process
involving the least baryon number change has change in Chern-Simons
charge of 1 unit.  Because of the factor
of $\alpha_W$ in front of $FF^d$ in the expression for the topological
charge, the fields strength with such integer values of the change
in Chern-Simons charge are therefore of order $1/g$.  The contribution
to the path integral is therefore of order
\be
	\int [dA]~ e^{-S[A]} \sim e^{-k/\alpha_W}
\ee
where $k$ is some constant.

The contributions to the path integral which generate
$B+L$ violation can be evaluated in weak coupling and are the
instanton solutions for SU(2) Yang-Mills theory.\cite{belavin}
Long ago,
't Hooft computed their contribution to $B+L$ violation and found
that the amplitude for this process is of order\cite{thooft}
\be
	\mid A_{B+L} \mid^2 \sim e^{-4\pi /\alpha_W} \sim 10^{-170}
\ee
This formula is valid for elementary processes which involve only
a few $W$ and $Z$ bosons.  The result is so small that it would
never be of any practical importance.

An amusing feature of the instanton amplitude is that it is
a finite action solution of the Euclidean equations of motion.
The Euclidean equations of motion differ from those of Minkowski
space only by the sign of the potential energy.  In order to get
a finite action solution, one must therefore find a solution which
goes from the top of a hill of the inverted potential to the top of another
hill.  This can only happen if the original potential had degenerate
minima.  The instanton solution takes the field between these
different minima.  It is straightforward to show that the action
associated with the instanton action is nothing more than the
WKB suppression factor associated with tunneling through a barrier
The tunneling is done at zero energy.  ' t Hooft's estimate of the
rate is therefore just a WKB tunneling computation of the rate
of B+L violation for processes which have an energy small compared to
the height of the barrier.  The barrier just separates sectors of the
theory with different values of the Chern-Simons charge.  We have
seen that these different sectors are degenerate in energy.  They
may be related to one another by gauge transformations which are
topologically non-trivial.  This follows because a topologically trivial
gauge transformation would allow the different sectors of the theory
to be joined together by a small field, yet we know that to get
an integer value of the Chern-Simons charge requires a field
of strength $1/g$.

The existence of energy barriers which separate the different topological
sectors of the theory tell us that the picture of baryon number violation
at energies of the order of or higher than that of the top of
the barrier must be much different than that which describes low
energy processes.  Naively, if we have energy higher than that of the barrier,
we expect that there exist classical non-forbidden processes which
allow the change of B+L. Before we discuss such processes,
let us first discuss some properties of the barrier.

Manton and Klinkhammer proposed a simple way to compute the energy
barrier in electroweak theory.\cite{manton}
If there is an energy barrier in ordinary
classical mechanics, a particle placed at the top of such a barrier
will not move.  If it is perturbed it will begin to roll down the hill.
There is therefore a static classical solution of the equations
of motion which is unstable under small perturbations which describes
the particle sitting at the top of the barrier.  Such a solution is
called a sphaleron, and its energy is the energy of the barrier.

The sphaleron solution of electroweak theory has been found,
and the energy of the barrier computed.  Because such a solution
is classical, it will have many quanta of order $1/\alpha$ in it.
Its size will be given by a typical electroweak size scale $M_W$.
The energy is of the form
\be
	E = A {{2M_W} \over \alpha_W}
\ee
where $A$ is a quantity which depends on the ratio of Higgs mass
to W boson mass.  It is of order 1 for all values of
$M_H/M_W$, and was computed by Manton and
Klinkhammer.\cite{manton}.  It is of the order of $10~ TeV$, so
that one can probe this energy scale in cosmology at a relatively
low temperature scale, and in SSC experiments.

There are two processes of interest at energies higher or of the order
of the height of the barrier.  There are finite temperature transition
processes relevant for cosmological processes such as baryon number
violation.\cite{krs} - \cite{arnold}
There are high energy collision processes where the particles in the initial
state have energies higher than that of the top of the
barrier.\cite{ringwald}-\cite{mvv}

On the other hand, the argument we constructed above that such processes
are necessarily of order $e^{-k/\alpha_W}$ seems to be convincing.
We found it was necessary to change Chern-Simons charge.  On the other
hand the action for any process is bounded from below by the action
for a single instanton.  To see this recall that the Higgs
part of the electroweak action is positive definite, so for
purpose of setting a lower bound on the action, we can ignore its contribution.
Using the identity that
\be
	(F\pm F^d)^2 = 2(F^2\pm FF^d)
\ee
shows that
\be
	\int F^2 \ge  ~\mid \int FF^d \mid
\ee
For the instanton $F = \pm F^d$ so the lower bound is saturated.
The contribution of sectors with topological charge is always
$\le e^{-2\pi /\alpha_W}$

On the other hand, when one computes the rate for baryon number
violation at high temperature, the rate is not exponentially suppressed
in contradictions with the above arguments.  How can this be?  The
evasion of the above argument is a consequence of coherent multiparticle
emission.  To understand this recall that the instanton field
is of order $1/g$.  The rate for instanton induced processes involving
$N$ fields is therefore of order
\be
	\mid A \mid^2 & \sim &  (1/\alpha_W)^N e^{-4\pi /\alpha_W}
\nonumber \\
                & \sim & e^{-4\pi /\alpha_W - Nln(\alpha_W) }
\ee
so that for large enough $N \sim 1/\alpha_W$ the enhancement due to the factors
of $1/\alpha_W$ due to the external vector boson lines begin to
dominate.

In general, this suggest that non-perturbative phenomena may become
important at high energy.  Suppose we consider multiparticle emission
in weak coupling.  For each particle emitted, there is a factor
of $\alpha_W$  On the other hand, coherence requires that there
is a factor of $N$ for each external line since we are assuming
the field of each external line adds coherently to that
of all the other particle involved in the process.  We have
\be
	\mid A \mid^2 \sim e^{Nln(N\alpha_W )}
\ee
so that when $N \sim 1/\alpha_W$ non-perturbative multiparticle
production may become big.

\section{Formulating the Problem}

In the previous section, we saw that non-perturbative many particle
processes in weakly coupled theories can become large.  The
example of finite temperature electroweak theory provides an example
where this is the case.

This does not necessarily imply that high energy
processes are also large.  If we consider a finite temperature process,
we can have a large number of particles in the initial state going into
a large number in the final state.  The total energy can be large because
there are a large number of particles.  Each particle need only
carry a small amount of energy.  It is this process which is large
at high temperature.

For high energy, there are only two particles in the initial state,
and they therefore must carry a high energy.  We might expect that
such processes would be suppressed by form factors.  The surprise is that
when such processes are computed to lowest order in weak coupling,
there are no form factors, and the largeness of the finite temperature
amplitudes would seem to imply that the high energy amplitudes are
also large.

To understand how this works, we repeat the analysis of
Ringwald.\cite{ringwald}-\cite{mvv}  To compute an instanton induced amplitude
to lowest order in weak coupling for scalar fields, we have
\be
	<\phi (p_1) \cdots \phi (p_N) > = \phi_{inst} (p_1) \cdots
\phi_{inst} (p_N)
\ee
Upon going to the residue of the pole at $p_i^2 = - m^2$, we
get point like amplitudes.  The amplitude has no form factor.
Analyzing vector amplitudes is more complicated but the same conclusion
holds.\cite{zakharov}

Therefore if the amplitudes for high energy B+L violation are not
large at high energy, then the weak coupling expansion must fail.
This is not too surprising since we are dealing with processes involving
$N \sim 1/\alpha_W$ particles, and we would naively expect that perturbation
theory breaks down.  Perturbation expansions are only expected to
be asymptotic expansions valid if the order of the expansion N is
$N << 1/\alpha_W$

It is amusing how the expansion breaks down.  Final state interactions
among soft particles are of order $\alpha_W$, but there are $N^2$ ways
to join lines together.  The final state corrections are of order
$\alpha_W N^2  \sim 1/\alpha_W$.  The initial final interactions
involve only $N$ particles, so we would naively expect these corrections
to be of order 1.  Mueller showed that there was an additional factor
proportional to the energy of the hard particle times that of the soft,
which again makes hard soft processes of order
$E_1E_2 \sim 1/\alpha_W$.\cite{mueller}
The hard-hard scattering process is naively of order $\alpha_W$, but the
Muller factor is $E_1 E_2 \sim 1/\alpha_W^2$ so that the overall factor
is $1/\alpha_W$.  Therefore all the corrections are large,
and of the same order of magnitude.

It can be proven that at least for the soft-soft interactions on this
amplitude, and presumably for the hard-soft and hard-hard interactions,
the corrections to the lowest order process exponentiate and the amplitude
is of the cross section for $B+L$ violation is of the form\cite{mattis} -
\cite{krt}
\be
	\sigma \sim exp(-4 \pi F(E/E_{sph}) /\alpha_W)
\ee
where $E_{sph}$ is the sphaleron energy of electroweak theory.
The weak coupling expansion resummed in this way is equivalent to a low
energy expansion valid when $E << E_{sph}$  The issue of whether
or not the rate for $B+L$ violation is large at high energy is the
issue of whether or not F has a zero.  This issue is clearly outside
the scope of naive weak coupling expansions.

\section{A Possible Solution of the Problem}

Although it is true that the naive weak coupling expansion has
broken down, it is not clear that there is no well defined semi-classical
approximation to the computation of scattering amplitudes for large numbers
of particles.  For any such approximation to be sensible, we must
of course require that the corrections to this approximation be small.

Using the example of a 1 dimensional integral, we can see that such a
semi-classical approximation may exist.  Consider the integral
\be
	I = \int dx~ x^N~e^{-(x^2-v^2)^2/g^2}
\ee
We will consider the case where $g << 1$ and $N >> 1$.  The naive weak coupling
expansion would correspond to doing a stationary phase approximation
on the exponential, that is, expanding around $x = \pm v$.  This is
a good approximation so long as $N \sim 1$.  For $N >> 1$, the proper way to
do the stationary phase approximation is to exponentiate the
power of x and extremize
\be
	-(x^2-v^2)^2/g^2 + Nln(x)
\ee
It is easy to check that this extremization yields a good approximation
to the integral for large N.

This suggests what may be done for field theory,\cite{mmy}
that is we write an expression for a Green's function as
\be
	\int [d\phi]~ \phi (p_1) \cdots \phi(p_N)~ e^{iS}
 = \int [d\phi]~ exp{\{ iS +ln(\phi(p_1) \cdots \phi(p_N))\} }
\ee
We are forced to write this expression in Minkowski space for as we will
soon see, the equations of motion we will derive will not
be Wick rotatable into Euclidean space.  The reason for this is simple
to understand:  the equations of motion including the exponentiated
external lines describe not vacuum to vacuum transitions, but
configurations with particles in the initial and final states.  In
this case, there will be anti-Feynman pieces to the boundary
conditions, and such terms may not be Wick rotated.

The equation of motion resulting by varying the above effective action are
\be
	-i{{\delta S} \over {\delta \phi (x)}} = \sum_i~e^{ip_ix}/\phi(p_i)
\ee
As we approach the residue of the pole, $p_i^2 = -m^2$, we see that the
right hand side of the equation vanishes.  On the other hand,
in the asymptotic region for large $x$, such a term would be coupled to
the free wave equation which also vanishes,
\be
	-i(-\partial^2 + m^2)\phi (x) = (-\partial^2 + m^2)
\sum_i~e^{ip_ix}/\phi_{res}(p_i)
\ee
where $\phi_{res}$ is the residue of the pole of the Fourier transformed field.
We see therefore that these source terms act as a boundary condition
for the fields requiring that asymptotically
\be
	lim_{t \rightarrow \pm \infty} \phi (x) = i\sum_i~e^{ip_ix}/\phi_{res}(p_i)
\ee
That is, we have anti-feynman boundary conditions for particles in the
initial state with a strength which must be determined self-consistently
by knowing the residue of the pole of the Fourier transformed
field.\cite{brown}  The sign of the limit in time must be
determined by the sign of the energy $E_i$, that is the sum on
the right hand side of the previous equations runs only over negative
energies for positive times and vice versa.

A remarkable feature of the above equations is that one can show that at least
to three loops expanding around the above solution, the corrections are small.
If we take the instanton solution and expand around it, we generate
the Mueller correction.

The problem with the above equation is that it is difficult to solve.
Although we started with a real field, we see that the
solution of the above equation with the anti-Feynman boundary conditions
must in general be complex.  Although the solution is complex,
there is still a conserved energy function. Because this energy
functional is no longer positive definite since the field is complex,
one can take a field which has a singularity, but due to cancellations
among the various terms in the energy functional the energy is conserved.
The general solution of the above problem therefore has singularities.
Although it may be possible to find a contour in the complex time
plane which avoids the singularity, one does not know the nature,
position or number of the singularities
until the equations have been solved.\cite{tinyakov}

These singularities lead to an amusing phenomenon.  Near a singularity,
a smooth field may be converted into a rapidly oscillating one.  This is
precisely what must happen in a high energy collision where
high energy particles produce many soft quanta.

It would seem to be a straightforward task to numerically integrate
the above equations.  It is not.  Closed boundary value problems
with singularities at internal points are difficult to solve when
the position nature and number of the singularities are
known.  Here we do not know
the position, number, and in some cases the nature of the
singularity.  In addition, one must do a calculation to very
high accuracy to handle the spectrum of Fourier modes needed.

Although the treatment used above was tailored for computing
scattering amplitudes, one could also write down expressions for
the cross section and try to evaluate them in a semi-classical
approximation.\cite{krt},\cite{tinyakov}.  This leads in the end to a set of
equations similar to the above, which might or might not
be simpler to analyze.  The problems with the singularities
and the complexity of the fields is however generic to
both problems, and their resolution demands solving the complicated
numerical problem outlined above, a problem which is difficult and
may prove impossible to solve to the needed precision.

\section{Acknowledgments}
I gratefully acknowledge countless conversations with Peter Arnold,
Michael Mattis, Valery Rubakov, Misha Shifman, Peter Tinyakov,
Arkady Vainshtein and Misha Voloshin on the problem discussed above.
I acknowledge support
under DOE  High Energy DE-AC02-83ER40105 and DOE Nuclear DE-FG02-87ER-40328.

\end{document}